\documentclass[12pt,preprint]{aastex}

\usepackage{epsfig}
\usepackage{lscape}
\usepackage{apjfonts}
\usepackage{mathptmx}

\newcommand{\Av}{A_V}

\newcommand{\dechms}[4]{$#1^{\rm h}#2^{\rm m}#3\mbox{$^{\rm s}\mskip-7.6mu.\,$}#4$}

\newcommand{\decdms}[4]{$#1^{\circ}#2'#3\mbox{$''\mskip-7.6mu.\,$}#4$}
\newcommand{\msec}[2]{$#1\mbox{$''\mskip-7.6mu.\,$}#2$}
\newcommand{\mmsec}[2]{$#1\mbox{$^s\mskip-7.6mu.\,$}#2$}
\newcommand{\mdeg}[2]{$#1\mbox{$^\circ \mskip-7.6mu.\,$}#2$}

\newcommand{\Lsun}{L$_{\odot}$}
\newcommand{\Msun}{M$_{\odot}$}

\begin{document}

\title{VLBA determination of the distance to nearby star-forming regions\\
       III. HP~Tau/G2 and the three-dimensional structure of Taurus}

\author{Rosa M. \ Torres, Laurent Loinard}
\affil{Centro de Radiostronom\'{\i}a y Astrof\'{\i}sica, 
       Universidad Nacional Aut\'onoma de M\'exico,\\
       Apartado Postal 72--3 (Xangari), 58089 Morelia, Michoac\'an, M\'exico;\\
       r.torres@astrosmo.unam.mx}

\author{Amy J.\ Mioduszewski}
\affil{National
    Radio Astronomy Observatory, Array Operations Center,\\ 1003
    Lopezville Road, Socorro, NM 87801, USA}

\and

\author{Luis F.\ Rodr\'{\i}guez}
\affil{Centro de Radiostronom\'{\i}a y Astrof\'{\i}sica,
       Universidad Nacional Aut\'onoma de M\'exico,\\
       Apartado Postal 72--3 (Xangari), 58089 Morelia, Michoac\'an, M\'exico}

\begin{abstract} 
  Using multi-epoch Very Long Baseline Array observations, we have
  measured the trigonometric parallax of the weak-line T Tauri star
  HP~Tau/G2 in Taurus. The best fit yields a distance of 161.2 $\pm$
  0.9 pc, suggesting that the eastern portion of Taurus (where HP
  Tau/G2 is located) corresponds to the far side of the complex.
  Previous VLBA observations have shown that T Tau, to the South of
  the complex, is at an intermediate distance of about 147 pc, whereas
  the region around L1495 corresponds to the near side at roughly 130
  pc. Our observations of only four sources are still too coarse to
  enable a reliable determination of the three-dimensional structure
  of the entire Taurus star-forming complex. They do demonstrate,
  however, that VLBA observations of multiple sources in a given
  star-forming region have the potential not only to provide a very
  accurate estimate of its mean distance, but also to reveal its
  internal structure. The proper motion measurements obtained
  simultaneously with the parallax allowed us to study the kinematics
  of the young stars in Taurus. Combining the four observations
  available so far, we estimate the peculiar velocity of Taurus to be
  about 10.6 km s$^{-1}$ almost completely in a direction  parallel to 
  the Galactic plane. Using our improved distance measurement, we
  have refined the determination of the position on the HR diagram
  of HP Tau/G2, and of two other members of the HP Tau group (HP 
  Tau itself and HP Tau/G3). Most pre-main sequence evolutionary 
  models predict significantly discrepant ages (by 5 Myr) for those 
  three stars --expected to be coeval. Only in the models of Palla \& 
  Stahler (1999) do they fall on a single isochrone (at 3 Myr).
\end{abstract}


\section{Introduction}

Several recent observations (e.g.\ Loinard et al.\ 2005, 2007, 2008;
Torres et al.\ 2007; Menten et al.\ 2007; Xu et al.\ 2006) have
demonstrated that multi-epoch VLBI observations can be used to measure
the trigonometric parallax of nearby young stars to better than a few
percent. Since the indirect methods traditionally used to estimate
the distance to nearby star-forming regions (e.g.\ Elias 1978a,b,
Kenyon et al.\ 1994, Knude \& Hog 1998) typically have uncertainties
of 20\%, VLBI observations have the potential of dramatically
improving our knowledge of the space distribution of star-formation
around the Sun. With this goal in mind, we have initiated a large
project aimed at accurately measuring the trigonometric parallax of a
sample of nearby magnetically active young stars using the 10-element
Very Long Baseline Array (VLBA) of the National Radio Astronomy
Observatory (NRAO). In previous papers of this series, we have
reported the distance and proper motions of three young stars in
Taurus (T~Tauri --Loinard et al.\ 2007, Hubble~4 and HDE~283572
--Torres et al.\ 2007). In the present article, we will concentrate on
HP~Tau/G2, a young star located near the eastern edge of the Taurus
complex.

The well-known variable star HP Tau was discovered by Cohen \& Kuhi
(1979) to be surrounded by a small group of young stars (called HP
Tau/G1, HP Tau/G2, and HP Tau/G3). HP Tau/G1 is located about 20$''$
north of HP Tau, whereas HP Tau/G2 and HP Tau/G3 are about 15$''$ to
its south-east (see the finding charts in Fig.\ 22 of Cohen \& Kuhi
1979). HP Tau/G2 and HP Tau/G3 are believed to form a gravitationally
bound system with a separation of about 10$''$.  Recently, HP Tau/G3
was itself found to be a tight binary (Richichi et al.\ 1994), so the
HP Tau/G2 - HP Tau/G3 system appears to be a hierarchical triple
system. HP~Tau/G2 is a weak-line T Tauri star of spectral type G0,
with an effective temperature of 6030 K (Briceño et al.\ 2002). It is
somewhat obscured ($\Av$ $\sim$ 2.1 mag) and has a bolometric
luminosity of 6.5 \Lsun\ (Briceño et al.\ 2002; Kenyon \& Hartmann
1995). This corresponds to an age of about 10.5 Myr and a mass of 1.58
\Msun\ (Brice\~no et al.\ 2002). The first radio detection of HP
Tau/G2 was reported by Bieging et al.\ (1984) who found a 5 GHz flux
of 5--7 mJy. A few years later, however, the flux had fallen to only
about 0.3 mJy (Cohen \& Bieging 1986). Such strong variability is
suggestive of non-thermal processes (e.g.\ Feigelson \& Montmerle
1999). The successful detection of HP Tau/G2 in VLBI experiments (at
levels of 1 to 3 mJy) by Phillips et al.\ (1991) confirmed the
non-thermal origin of the radio emision.

\section{Observations and data calibration}

In this paper, we will make use of a series of nine continuum 3.6 cm
(8.42 GHz) observations of HP~Tau/G2 obtained between September 2005
and December 2007 with the VLBA (Tab.\ 1). Our phase center was at
$\alpha_{J2000.0}$ = \dechms{04}{35}{54}{161}, $\delta_{J2000.0}$ =
+\decdms{22}{54}{13}{492}. Each observation consisted of series of
cycles with two minutes spent on source, and one minute spent on the
main phase-referencing quasar J0426+2327, located \mdeg{2}{14}
away. J0426+2327 is a compact ICRF source (Ma et al.\ 1998) whose
absolute position is known to better than 0.7 mas.  Every 24 minutes,
we also observed two secondary calibrators (J0435+2532 and J0449+1754)
which, together with the primary calibrator, form a triangle around
the astronomical target (Fig.\ 1).

\begin{landscape}
\begin{deluxetable}{lccccccc}
\tablecaption{Measured source positions and flux densities}
\tablehead{
\colhead{Mean UT date}       &  
\colhead{Julian Day}         &
\colhead{$\alpha$ (J2000.0)} &
\colhead{$\sigma_\alpha$}    &
\colhead{$\delta$ (J2000.0)} &
\colhead{$\sigma_\delta$}    &
\colhead{$f_\nu$}            &
\colhead{$\sigma$}           \\%
~~(yyyy.mm.dd ~~ hh:mm)~~ & & \multicolumn{1}{c}{04$^h$35$^m$} & & \multicolumn{1}{c}{22$^\circ$54$'$} & & (mJy) & (mJy beam$^{-1}$)}%
\startdata
2005.09.07 ~~ 12:36 \dotfill & 2453621.02 & \mmsec{54}{1613574} & \mmsec{0}{0000034} & \msec{13}{41131} & \msec{0}{00009} & 0.71 & 0.06 \\%
2005.11.16 ~~ 08:01 \dotfill & 2453690.83 & \mmsec{54}{1612212} & \mmsec{0}{0000019} & \msec{13}{40798} & \msec{0}{00007} & 0.97 & 0.07 \\%
2006.01.23 ~~ 03:33 \dotfill & 2453758.65 & \mmsec{54}{1609360} & \mmsec{0}{0000042} & \msec{13}{40439} & \msec{0}{00010} & 0.99 & 0.07 \\%
2006.03.31 ~~ 23:06 \dotfill & 2453826.46 & \mmsec{54}{1610940} & \mmsec{0}{0000032} & \msec{13}{40161} & \msec{0}{00010} & 0.68 & 0.07 \\%
2006.06.10 ~~ 18:27 \dotfill & 2453897.27 & \mmsec{54}{1617432} & \mmsec{0}{0000008} & \msec{13}{39959} & \msec{0}{00002} & 3.06 & 0.08 \\%
2006.09.08 ~~ 12:33 \dotfill & 2453987.02 & \mmsec{54}{1623557} & \mmsec{0}{0000024} & \msec{13}{39681} & \msec{0}{00007} & 1.08 & 0.06 \\%
2007.06.04 ~~ 18:56 \dotfill & 2454256.29 & \mmsec{54}{1627018} & \mmsec{0}{0000038} & \msec{13}{38267} & \msec{0}{00013} & 0.63 & 0.07 \\%
2007.09.03 ~~ 12:53 \dotfill & 2454347.04 & \mmsec{54}{1633509} & \mmsec{0}{0000025} & \msec{13}{38127} & \msec{0}{00008} & 0.78 & 0.06 \\%
2007.12.04 ~~ 06:51 \dotfill & 2454438.79 & \mmsec{54}{1631378} & \mmsec{0}{0000037} & \msec{13}{37605} & \msec{0}{00009} & 0.76 & 0.05 \\%
\enddata
\end{deluxetable}
\end{landscape}

\begin{figure}[!b]
\centerline{\includegraphics[height=0.45\textwidth,angle=-90]{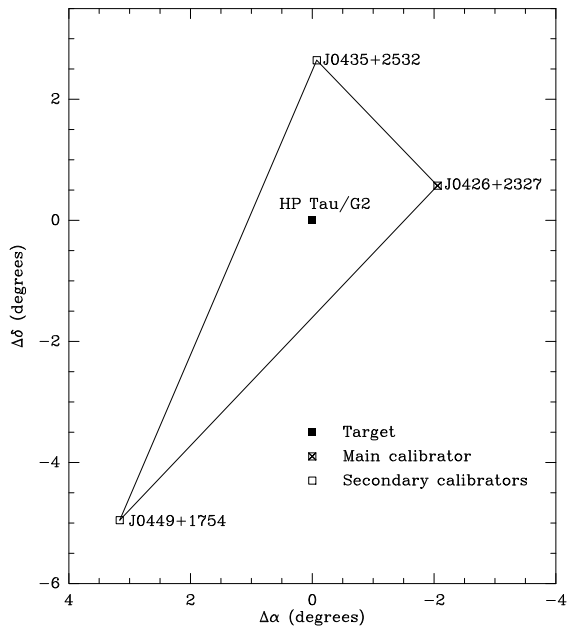}}
\caption{Relative position of the astronomical target, the main
calibrator (J0426+2327), and the secondary calibrators (J0435+2532,
and J0449+1754).}
\end{figure}

The data were edited and calibrated using the Astronomical Image
Processing System (AIPS --Greisen 2003). The basic data reduction
followed the standard VLBA procedure for phase-referenced
observations, and was described in detail in Loinard et al.\
(2007). Using the secondary calibrators, we applied the multi-source
calibration strategy described in Torres et al.\ (2007) to correct for
systematic errors due to inaccuracies in the troposphere model used,
and to clock, antenna, and source position errors. This resulted in
significant improvements in the final phase calibration and image
quality.

Six hours of telescope time were allocated to each of the first six
observations, whereas 9 hours were allocated for each of the last
three. Because of the time spent on the calibrators, however, only
about 3 and 5 hours were actually spent on source during the first six
and the following three observations, respectively. Once calibrated,
the visibilities were imaged with a pixel size of 50 $\mu$as after
weights intermediate between natural and uniform (ROBUST = 0 in AIPS)
were applied. This resulted in typical r.m.s.\ noise levels of
0.06--0.08 and 0.05--0.07 mJy beam$^{-1}$ during the first six and the
last three observations, respectively (Tab.\ 1). The source was
detected with a signal to noise ratio of 10 or better at each epoch
(Tab.\ 1). The source position (also listed in Tab.\ 1) was
determined using a 2D Gaussian fitting procedure (task JMFIT in
AIPS). This task provides an estimate of the position error (columns 3
and 5 of Tab.\ 1) based on the expected theoretical astrometric
precision of an interferometer (Condon 1997). However, in spite of the
extra calibration steps taken to improve the phase calibration,
uncorrected systematic errors still exist, and must be added
quadratically to the values listed in Tab.\ 1. These remaining
systematic errors are difficult to estimate {\it a priori}, and may
depend on the structure of the source under consideration. Here, we
will estimate them from the fits to the data (see below).

\section{Results}

\subsection{Variability and morphological changes}

The flux of HP Tau/G2 was fairly constant around 0.8 mJy at eight of
our nine observations. During the fifth observation (June 2006),
however, HP Tau/G2 clearly underwent a flaring event, reaching a flux
3 to 4 times higher than that at the other epochs (Fig.\ 2). This type
of variability is not unexpected for non-thermal sources such as HP
Tau/G2 (e.g.\ Feigelson \& Montmerle 1999; Loinard et al.\ 2008), and
is consistent with previous radio measurements (Bieging et al.\ 1984;
Cohen \& Bieging 1986; Phillips et al.\ 1991; see Sect.\ 1). The radio
source was found to be unresolved at all epochs except the 7th
(obtained in June 2007) when it was clearly extended in the
north-south direction, with a deconvolved size in that direction of
about 2.5 mas (Fig.\ 3). This increase in the source size might be due to
changes in the structure of the active magnetosphere of HP
Tau/G2. Interestingly, however, this observation does not correspond
to an epoch when the source was particularly bright, nor particularly
dim. In any event, the determination of the source position for that
epoch is adversely affected by the fact that the source is extended
(see below).

\begin{figure}[!t]
\centerline{\includegraphics[width=.4\textwidth,angle=270]{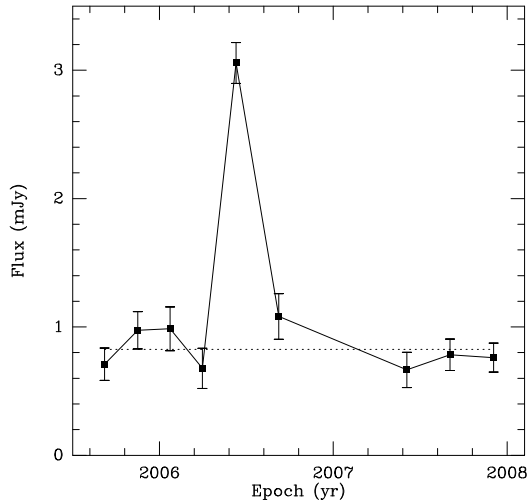}}
\caption{Time evolution of the 3.6 cm flux of HP Tau/G2. Note the 
flare during the fifth observation.}
\end{figure}

\begin{figure}[!t]
\centerline{\includegraphics[width=.4\textwidth,angle=270]{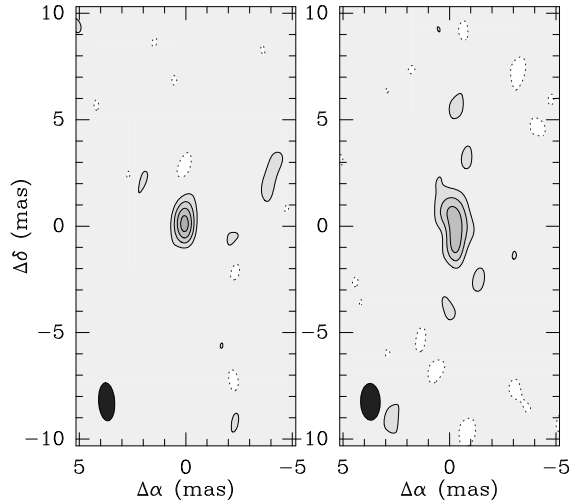}}
\caption{Images of HP Tau/G2 at the sixth (left) and seventh (right)
epochs. The first contour and the contour spacing in both images is 
0.12 mJy beam$^{-1}$, and the synthesized beams are shown 
at the bottom left of each panel. Note how the source is elongated in 
the north-south direction during the seventh epoch, whereas it is
unresolved during the sixth observation.}
\end{figure}

\subsection{Astrometry}

The displacement of HP Tau/G2 on the celestial sphere is the
combination of its trigonometric parallax ($\pi$) and proper motion
($\mu$). Since HP Tau/G2 is a member of a triple system (see Sect.\
1), we should in principle describe its proper motion as the
combination of the uniform motion of the center of mass and a
Keplerian orbit. This is not necessary, however, because the orbital
period of the system must be very much longer than the timespan
covered by our observations. If we assume that the total mass of the
HP Tau/G2 -- HP Tau/G3 system is 2--3 \Msun\ and that the current
observed separation is a good estimate of the system's semi-major
axis, then the orbital period is expected to be 35,000 to 45,000
yr. This is indeed very much longer that the 2 yr covered by our
observations, and the acceleration terms can be safely ignored.  The
astrometric parameters were calculated using the SVD-decomposition
fitting scheme described by Loinard et al.\ (2007). The necessary
barycentric coordinates of the Earth, as well as the Julian date of
each observation were calculated using the Multi-year Interactive
Computer Almanac (MICA) distributed as a CDROM by the US Naval
Observatory. The reference epoch was taken at the mean of our
observations: JD 2454029.90 $\equiv$ J2006.81.

Since the source was elongated in the north-south direction during the
seventh observation, two different fits were made: one where the
seventh epoch was included, and one where it was ignored. When the
seventh epoch is included, we obtain the following astrometric
parameters:

\begin{eqnarray}
\alpha_{J2006.81} & = & \mbox{ \dechms{04}{35}{54}{162033} } ~ \pm ~ \mbox{ \mmsec{0}{000003} } \nonumber \\%
\delta_{J2006.81} & = & \mbox{ \decdms{22}{54}{13}{49345} } ~ \pm ~ \mbox{ \msec{0}{000020} } \nonumber \\%
\mu_\alpha \cos \delta& = & 13.90 ~ \pm ~ 0.06 ~ \mbox{mas yr$^{-1}$} \nonumber \\%
\mu_\delta & = & -15.6 ~ \pm ~ 0.3 ~ \mbox{mas yr$^{-1}$} \nonumber \\%
\pi & = & 6.19 ~ \pm ~ 0.07 ~ \mbox{mas.} \nonumber
\end{eqnarray}

\noindent This corresponds to a distance of 161.6 $\pm$ 1.7 pc. The
postfit r.m.s.\ in this case is 0.12 mas in right ascension and 0.51
mas in declination\footnote{The residual is much larger in declination
  than in right ascension. We will come back to this point in Sect.\
  3.3.}. To obtain a reduced $\chi^2$ of one in both right ascension
and declination, one must add quadratically 8.8 $\mu$s and 0.59 mas in
right ascension and declination, respectively, to the errors listed in
Tab.\ 1. The uncertainties on the parameters quoted above include
these systematic contributions. Note that the seventh epoch
contributes significantly to the total post-fit r.m.s.\ since the
position corresponding to that observation is farther from the fit
(both in right ascension and declination) than that at any other epoch
(Fig.\ 4). If the seventh observation is ignored, the best fit yields
the following parameters:

\begin{eqnarray}
\alpha_{J2006.81} & = & \mbox{ \dechms{04}{35}{54}{162030} } ~ \pm ~ \mbox{ \mmsec{0}{000002} } \nonumber \\%
\delta_{J2006.81} & = & \mbox{ \decdms{22}{54}{13}{49362} } ~ \pm ~ \mbox{ \msec{0}{000014} } \nonumber \\%
\mu_\alpha \cos \delta& = & 13.85 ~ \pm ~ 0.03 ~ \mbox{mas yr$^{-1}$} \nonumber \\%
\mu_\delta & = & -15.4 ~ \pm ~ 0.2 ~ \mbox{mas yr$^{-1}$} \nonumber \\%
\pi & = & 6.20 ~ \pm ~ 0.03 ~ \mbox{mas.} \nonumber
\end{eqnarray}

\noindent All these parameters are consistent within 1$\sigma$ with
those obtained when the seventh observation is included. The
corresponding distance in this case is 161.2 $\pm$ 0.9 pc, and the
post-fit r.m.s.\ is 0.058 mas in right ascension and 0.33 mas in
declination, significantly better than in the previous fit.  Indeed,
to obtain a reduced $\chi^2$ of one in both right ascension and
declination, one must only add quadratically 3.65 $\mu$s and 0.38 mas
to the formal errors delivered by JMFIT. Again, the uncertainties on
the parameters quoted above include these systematic
contributions. 

As mentioned earlier, the source during the seventh epoch was extended,
and the astrometry consequently less reliable. Since the fit when it
is ignored is clearly much better than that when it is included, we
consider the second fit above our best result.

\begin{figure}[!t]
\centerline{\includegraphics[width=.47\textwidth,angle=270]{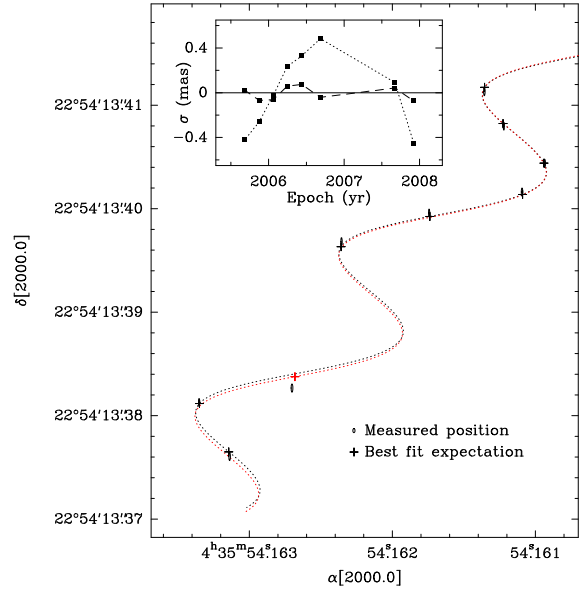}}
\caption{Measured positions and best fits for HP~Tau/G2. The observed
  positions are shown as ellipses, the size of which represents the
  error bars. Two fits are shown: the dotted black line corresponds to
  the fit where the 7th epoch is ignored, whereas the dotted red line
  is the fit where it is included. Note that the 7th observation falls
  significantly to the south of either fit. The inset shows the fit
  residuals (of the fit without the 7th epoch) in right ascension (dashed line) 
  and declination (dotted line). Note the large residuals in declination.}
\end{figure}

\begin{figure}[!t]
\centerline{\includegraphics[width=.38\textwidth,angle=270]{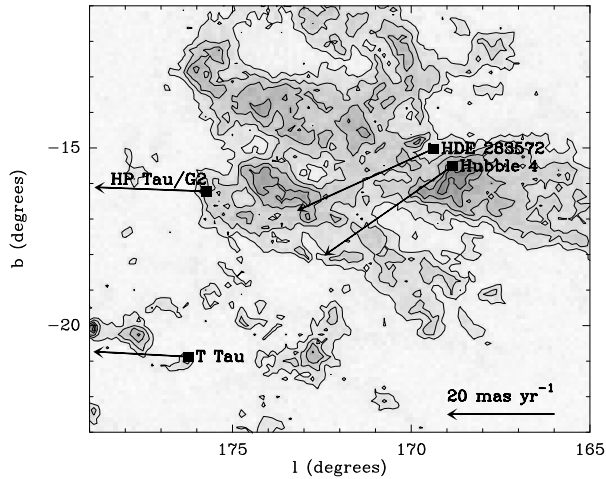}}
\caption{Positions and proper motions of Hubble 4, HDE~283572, T Tau,
and HP Tau/G2 superposed on the CO(1-0) map of Taurus from Dame et al.\
(2001).}  
\end{figure}

\subsection{Error analysis}

It is noteworthy that, whether or not the seventh epoch is included,
the post-fit r.m.s.\ and the systematic error contribution that must
be added to the uncertainties quoted in Tab.\ 1, are much larger in
declination than in right ascension. Fortunately, this large
declination contribution does not strongly affect the distance
determination, because the strongest constraints on the parallax come
from the right ascension measurements. Interestingly, this is, with
Hubble 4 (Torres et al.\ 2007), the second source for which we find
large systematic declination residuals. Astrometric fitting of
phase-referenced VLBI observations is usually worse in declination
than in right ascension (e.g.\ Fig.\ 1 in Chatterjee et al.\ 2004) as
a result of residual zenith phase delay errors (Reid et al.\ 1999). In
the case of Hubble 4, however, we argued that the large post-fit
declination r.m.s.\ might trace the reflex motion caused by an unseen
companion, because a periodicity of about 1.2 yr could be discerned in
the residuals.  In the present source, the case for a periodicity is
less clear (Fig.\  4, inset), but the residuals are clearly not
random. Interestingly, the large residuals are in the same north-south
direction as the extension of the source seen during our 7th
observation. This orientation might, therefore, correspond to a
preferred direction of the system along which it tends to vary more
strongly. Additional observations will clearly be necessary to
settle this issue.

\begin{landscape}
\begin{table*}
\caption{Radial velocities, proper motions, heliocentric, and peculiar velocities  in Galactic coordinates for the 4 sources 
observed with the VLBA so far.}
\begin{tabular}{lcrrccccccc}
\hline
\multicolumn{1}{c}{Source} & \multicolumn{1}{c}{$V_r$} &  \multicolumn{1}{c}{$\mu_\ell$cos$(b)$} &  
\multicolumn{1}{c}{$\mu_b$} & $U$ & $V$ & $W$ & $u$ & $v$ & $w$ &References\tablenotemark{a}\\%
 & (km s$^{-1}$) & \multicolumn{2}{c}{(mas yr$^{-1}$)} & \multicolumn{3}{c}{(km s$^{-1}$)} & 
\multicolumn{3}{c}{(km s$^{-1}$)}\\%
\hline
HP~Tau/G2                   & 17.7 $\pm$ 1.8 & $+$20.90 $\pm$ 0.07 &  $+$0.82 $\pm$ 0.10  & $-$18.59 & $-$14.65 & $-$4.50  & $-$8.59 & $-$9.40 & $+$2.67 & 1,2 \\%
Hubble~4                    & 15.0 $\pm$ 1.7 & $+$23.94 $\pm$ 0.12 & $-$16.74 $\pm$ 0.15  & $-$14.96 & $-$12.66 & $-$14.30 & $-$4.96 & $-$7.41 & $-$7.13 & 3,4 \\%
HDE~283572                  & 15.0 $\pm$ 1.5 & $+$25.53 $\pm$ 0.05 & $-$11.61 $\pm$ 0.06  & $-$15.89 & $-$13.07 & $-$10.84 & $-$5.89 & $-$7.82 & $-$3.67 & 3,2\\%
T~Tau\tablenotemark{b}      & 19.1 $\pm$ 1.2 & $+$17.76 $\pm$ 0.03 &  $+$0.99 $\pm$ 0.04  & $-$19.09 & $-$11.27 & $-$6.30  & $-$9.09 & $-$6.02 & $+$0.87 & 4,5\\%
\hline
\end{tabular}
\tablenotetext{a}{1=This work; 2=Walter et al.\ 1988; 3=Torres et al.\ 2007; 4=Hartmann et al.\ 1986; 5=Loinard et al.\ 2007}
\tablenotetext{b}{The radial velocity and proper motions used here are those of T Tau N. The radial velocities for T Tau Sa and T Tau Sb are available in Duch\^ene et al.\ (2002) and are very similar.
\vspace{1cm}}
\end{table*}
\end{landscape}

\section{Discussion}

\subsection{Kinematics of the sources in Taurus}

For Galactic sources, it is interesting to express the proper motions
in Galactic coordinates rather than in the equatorial system naturally
delivered by the VLBA. The results for HP Tau, and the three sources
previously observed with the VLBA are given in columns 3 and 4 of
Tab.\ 2. Interestingly, the proper motion of HP Tau/G2 is very similar
to that of T Tau, but significantly different from those of Hubble 4
and HDE~283572 (which are themselves very similar to each other).  HP
Tau/G2 and T Tau also happen to both be located on the eastern side of the
Taurus complex, whereas Hubble 4 and HDE~283572 are both around Lynds
1495 near the center of the complex (Fig.\ 5).

There is a fifth star in Taurus (V773 Tau) with VLBI-based proper
motion and trigonometric parallax measurements (Lestrade et al.\
1999). V773 Tau is located about a degree south-west of Hubble 4,
and the proper motions reported by Lestrade et al.\ (1999) are similar 
to those of Hubble 4 and HDE~283572  (see Fig.\ 4 in Torres et al.\ 
2007). It is now known that V773 Tau is a quadruple system composed 
of a tight spectroscopic binary orbited by two companions. The
radio source observed by Lestrade et al.\ (1999) is associated with 
the spectroscopic binary (the other two stars do not appear to be 
detectable radio emitters). In several VLBI observations (e.g.\
Phillips et al.\ 1991, Boden et al.\  2007, Torres et al.\  2008), the 
radio source has been reported to be double, with each component
tracing one of the stars in the spectroscopic binary. This binarity
was not taken into account by Lestrade et al.\ (1999), and may
have affected their parallax measurement. Indeed, Boden et al.
(2007) recently modeled the orbit of V773 Tau combining
spectroscopic observations, Keck Interferometer data and 
VLBA imaging. The distance to V773 Tau that they obtain
(136.2 $\pm$ 3.7 pc) is somewhat smaller than the value
(148.4$^{+5.7}_{-5.3}$ pc) reported by Lestrade et al.\ (1999).
Because of this slight discrepancy, we will not include V773 Tau in
the present analysis. It should be mentioned that we are currently
analyzing new multi-epoch VLBA observations of V773 Tau 
designed to constrain both its distance and orbital motions. 
These data will be published in a forthcoming paper.

Knowing the distance to the sources with high accuracy, it is possible
to transform the observed proper motions into transverse velocities.
Combining this information with radial (Heliocentric) velocities
taken from the literature (second column of Tab.\ 2), it becomes
possible to construct the three-dimensional velocity vectors. It is
common to express these vectors on a rectangular $(X,Y,Z)$ coordinate
system centered in the Sun, with $X$ pointing towards the Galactic
center, $Y$ in the direction of Galactic rotation, and $Z$ towards the
Galactic North Pole. In this system, the coordinates of the
Heliocentric velocities will be written $(U,V,W)$. As a final step, it
is also possible to calculate the peculiar velocity of the stars. This
involves two stages: first, the peculiar motion of the Sun must be
removed to transform the Heliocentric velocities into velocities
relative to the LSR. Following Dehnen \& Binney (1998), we will use
$u_0$ = +10.00 km s$^{-1}$, $v_0$ = +5.25 km s$^{-1}$, and $w_0$ =
+7.17 km s$^{-1}$ for the peculiar velocity of the Sun expressed in the
coordinate system defined above. The second stage consists in
estimating the difference in circular velocity between Taurus and the
Sun, so the peculiar velocities are expressed relative to the LSR
appropriate for Taurus, rather than relative to the LSR of the
Sun. This was done assuming the rotation curve of Brand \& Blitz
(1993), and represents a small correction of only about 0.3 km
s$^{-1}$. We will write $(u,v,w)$ the coordinates of the peculiar
velocity of the sources. Both $(U,V,W)$ and $(u,v,w)$ are given in
Tab.\ 2 for the four sources considered here. Their projections onto
the $(X,Y)$, $(X,Z)$, and $(Y,Z)$ planes are shown in Fig.\ 6.

The mean heliocentric velocity and the velocity dispersion of the four
sources are:

\begin{eqnarray}
U & = & -17.1 \pm 1.7 \mbox{~km s$^{-1}$} \\%
V & = & -12.9 \pm 1.2 \mbox{~km s$^{-1}$} \\%
W & = & -9.0 \pm 3.8 \mbox{~km s$^{-1}$.}
\end{eqnarray}

\noindent
These values are similar to those reported by Bertout \& Genova (2006)
for a larger sample of young stars in Taurus with optically measured
proper motions. Note that the velocity dispersion in the $W$ direction
is somewhat artificially high because (as noted earlier) Hubble 4 and
HDE~283572 on the one hand, and T Tau and HP Tau/G2 on the other,
clearly have different vertical velocities. They likely belong to two
different kinematic sub-groups.

The mean peculiar velocity of the four sources considered here is:

\begin{eqnarray}
u & = & -7.1 \pm 1.7 \mbox{~km s$^{-1}$} \\%
v & = & -7.7 \pm 1.2 \mbox{~km s$^{-1}$} \\%
w & = & -1.8 \pm 3.8 \mbox{~km s$^{-1}$.}
\end{eqnarray}

\noindent
We argue that this is a good estimate of the mean peculiar velocity of
the Taurus complex. This velocity is almost entirely in the $(X,Y)$
plane. Thus, although Taurus is located significantly out of the
mid-plane of the Galaxy (about 40 pc to its south), it appears to be
moving very little in the vertical direction. The motion in the
$(X,Y)$ plane, on the other hand, is fairly large, leading to a total
peculiar velocity $(u^2+v^2+w^2)^{0.5}$ $=$ 10.6 km s$^{-1}$. 
According to Stark \& Brand (1989), the one-dimensional velocity
dispersion of giant molecular clouds within 3 kpc of the Sun is about
8 km s$^{-1}$. As a consequence, each component of the peculiar 
velocity of a given molecular cloud is expected to be of that
order, and our determination of the mean peculiar velocity of 
Taurus is in reasonable agreement with that prediction. Another
useful comparison is with the velocity dispersion of young main 
sequence stars. For the bluest stars in their sample (corresponding
to early A stars), Dehnen \& Binney (1998) found velocity dispersions
of about 6 km s$^{-1}$ in the vertical direction, and of 10-14 km s$^{-1}$
in the $X$ and $Y$ directions. The young stars in Taurus are significantly
younger than typical main sequence early A stars, so one would expect
young stars in Taurus to have peculiar velocities somewhat smaller than 
6 km s$^{-1}$ in the vertical direction, and than 10--14 km s$^{-1}$ in the 
$X$ and $Y$ directions. This is indeed what is observed. Note, however,
that Taurus is not among the star-forming regions with the smallest
peculiar velocities. In Orion, G\'omez et al.\ (2005) found a difference between
expected and observed proper motions smaller than 0.5 km s$^{-1}$.

One last comment should be made here. The data presented here and in
the other papers of this series yield proper motions and trigonometric
parallaxes that, together, enable the measurement of transverse velocities with 
an accuracy of about 1\%. For sources in Taurus, this corresponds to
an absolute error better than 0.1 km s$^{-1}$. In comparison, the
radial velocity measurements available in the literature have typical
uncertainties of 1 to 2 km s$^{-1}$. To take full advantage of the
VLBA data, it will become important to measure radial velocities with
a significantly improved accuracy.

\begin{figure*}[!t]
\centerline{\includegraphics[width=.28\textwidth,angle=270]{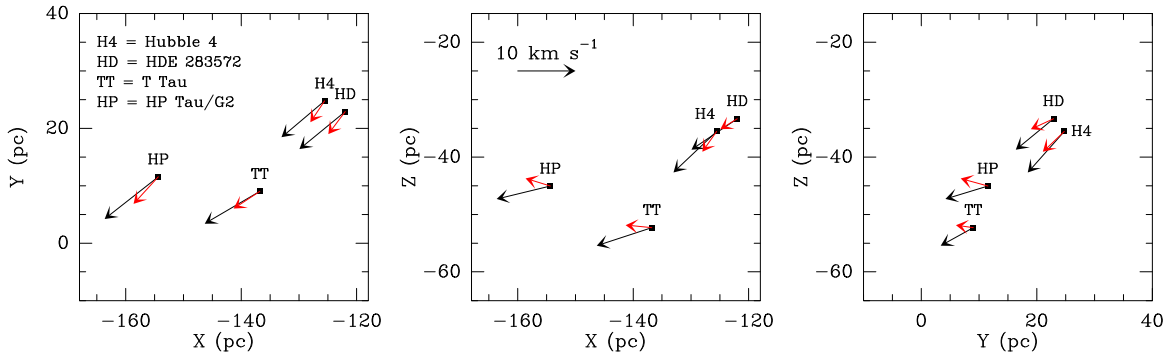}}
\caption{Heliocentric velocities (black arrows) and peculiar velocities
(red arrows) for the four stars in Taurus with VLBA-based distance
determinations.}  
\end{figure*}

\subsection{Distance and structure of the Taurus association}

Taking the mean of the four VLBA-based parallax measurements
available (HP Tau/G2, T Tau, Hubble 4 and HDE~283572), we can
estimate the mean parallax to the Taurus complex to be $\bar{\pi}$ =
7.08 mas. This corresponds to a mean distance $\bar{d}$ of 141.2 pc,
in good agreement with previous estimates (Kenyon et al.\ 1994).

The angular size of Taurus is about 10$^\circ$, corresponding to a
physical size of roughly 25 pc. It would be natural to expect that the
depth of Taurus might be similar, and that different sources may be
found at significantly different distances from us. The observations
presented here and in the previous papers of this series are, however,
the first ones with enough accuracy to directly probe the depth of the
Taurus complex. They reveal that HP Tau is about 30 pc farther than
Hubble 4 and HDE~283572, and that Taurus is at least as deep as it is
wide. A trivial but important consequence is that using the mean
distance indiscriminately for all the stars in the complex will result
in systematic errors at the levels of about 10\%. To reach higher
accuracy, one will have to reconstruct the complete three-dimensional
structure of Taurus. The number of sources considered so far is
obviously too limited to obtain such a complete view.  It is
interesting to note, however, that Hubble 4 and HDE~283572 which are
very near one another in projection, and share the same kinematics
(See Sect.\ 4.2), are also found to be at similar distances from us
($\sim$ 130 pc). This suggests that there exist in that region
(corresponding to the surroundings of the dark cloud Lynds 1495) a
coherent spatio-kinematical structure at about 130 pc. Observations
with an astrometric precision similar to that of the data presented
here for several dozen young stars would allow the identification of
several such coherent groups across the complex. This, in turn, would
allow a fairly accurate re-construction of a three-dimensional
structure of Taurus. Currently, Very Long Baseline Interferometry is
the only technique with sufficient accuracy to carry out the necessary
observations.

\begin{figure*}[!t]
\centerline{\includegraphics[width=.95\textwidth,angle=0]{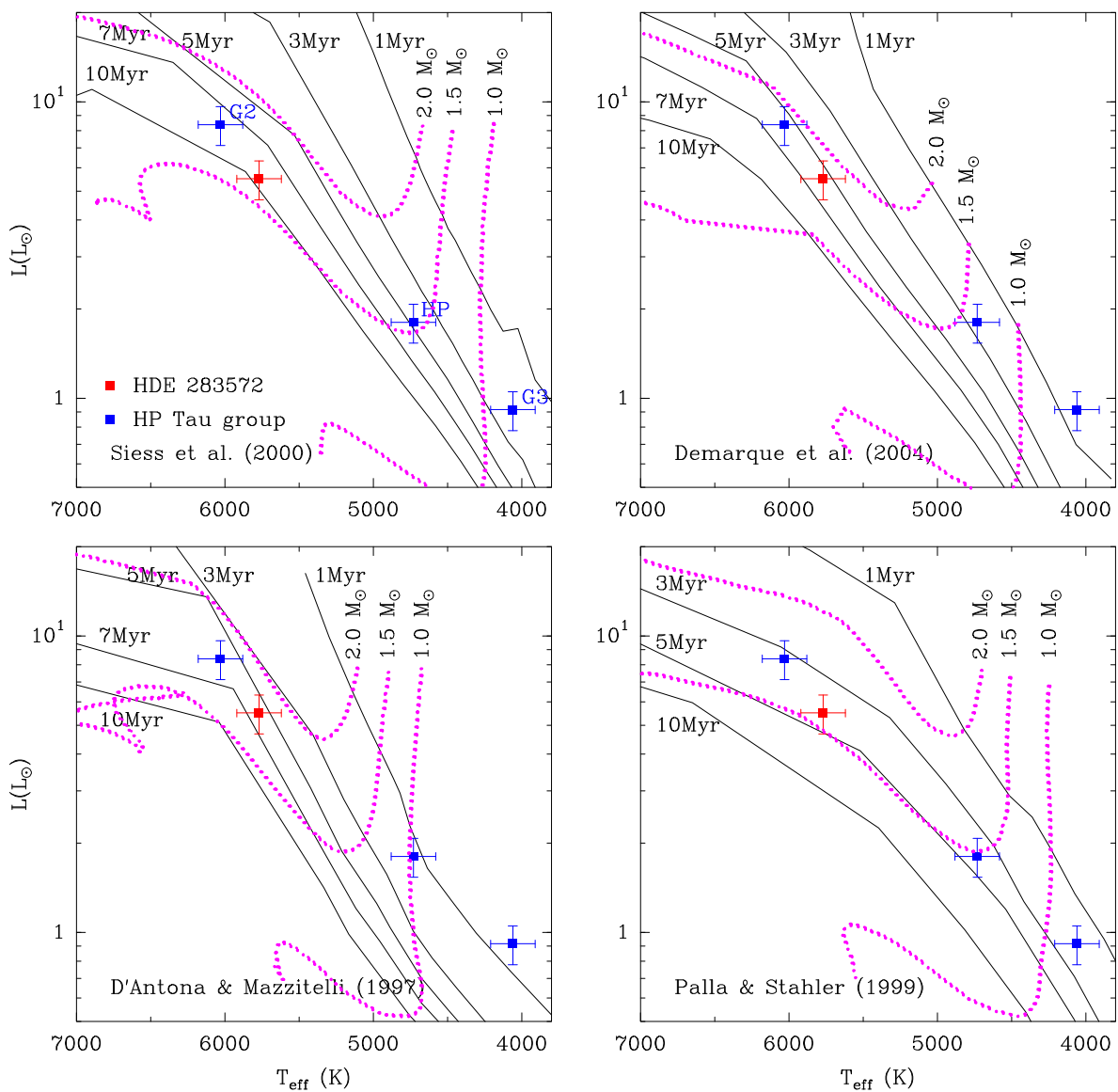}}
\caption{Positions of the three HP Tau members (blue symbols) and 
of HDE~283572 (red symbol) on an HR diagram. (From the coolest to
the warmest, the three stars in the HP Tau group are HP Tau/G3, HP Tau,
and HP Tau/G2, as indicated in the first panel.) Isochrones (full
back lines) are shown at 1, 3, 5, 7, and 10 Myr for various models.
For the same models, evolutionary tracks for stars of 1.0, 1.5, and
2.0 \Msun\ are also shown as dotted magenta lines.}  
\end{figure*}

\subsection{Comparison with theoretical evolutionary tracks}

As mentioned in Sect.\ 1, HP Tau/G2 is a member of a compact 
group of four young stars, comprising HP Tau itself, HP Tau/G1, 
G2, and G3. Given the small angular separations between them, 
the members of this group are very likely to be physically associated
--indeed, HP Tau/G2 and G3 are thought to form a bound system. They
are, therefore, very likely to be at the same distance from the Sun.
Using our accurate estimate of the distance to HP Tau/G2, we are
now in a position to refine the determination of the luminosities of 
all four stars. Little is known about HP Tau/G1, but the effective 
temperature and the bolometric luminosity (obtained assuming 
$d$ = 142 pc) of the other three members are given in Brice\~no et 
al.\  (2002). Those values (corrected to the new distance) allow us 
to place the stars accurately on an HR diagram (Fig.\ 7). 

From their position on the HR diagram, one can (at least in
principle) derive the mass and age of the stars using theoretical 
pre-main sequence evolutionary codes. Several such models
are available, and we will use four of them here\footnote{The
models by Baraffe et al.\  (1998) will not be used because they
do not cover the mass range of our stars.}: those of
Siess et al.\ (2000), Demarque et al.\ (2004; known as the
Yonsei-Yale $Y^2$ models), D'Antona \& Mazzitelli (1997),
and Palla \& Stahler (1999). The isochrones for those four
models at 1, 3, 5, 7, and 10 Myr are shown as solid black lines
in Fig.\ 7. Also shown are the evolutionary tracks (from the
same models) for stars of 1.0, 1.5, and 2.0 \Msun. The three
HP Tau members are shown as blue symbols, and HDE~283572 
(from Torres et al.\ 2007) is shown as a red symbol. (We will
discuss momentarily the reason for incorporating that source in 
the present analysis.)

A number of interesting points can be seen from Fig.\ 7. First,
there is reasonable agreement (within 40\%, see below) between
the masses predicted by different models. The best case is that
of HP Tau/G2, for which the different models predict masses 
consistent with each other at the 10\% level (between 1.7 and 
1.9 \Msun). The situation for HP Tau is somewhat less favorable, 
since the models of Siess et al.\
(2000) or Palla \& Stahler (1999) predict a mass of $\sim$ 1.5 \Msun, 
whereas those of D'Antona \& Mazzitelli (1997) predicts a significantly 
smaller mass of $\sim$ 1.0 \Msun. Thus, there is a 35\% spread in 
the values predicted by different models for the mass of that source. 
The least favorable situation is for HP Tau/G3. The mass of 
that source is about 0.8 \Msun\ according to
the models of Siess et al.\ (2000), but slightly less than 0.5 \Msun\
according to those D'Antona \& Mazzitelli (1997). This is a 40\% 
discrepancy. This tendency for
 pre-main sequence evolutionary models to become more
 discrepant at lower mass had been noticed before, and is
 discussed at length in Hillenbrand et al.\ (2008). In the
 absence of dynamically measured masses, it is impossible
 to assess which of the models used here provides the
 ``best'' answer.
 
 Another interesting issue is related to the age predictions of
 the different models. Since the different members of the HP Tau
 group are likely to be physically associated, they are expected to
 be nearly coeval. This is particularly true of HP Tau/G2
 and HP Tau/G3 which are believed to form a loose binary 
 system. Interestingly, most models predict significantly different
 ages for the three sources (see Fig.\ 7). The models by Siess
 et al.\ (2000) predict ages of about 8 Myr and 3 Myr for HP Tau/G2 and
 HP Tau/G3, respectively. A similar 5 Myr age difference is found
 for the models of Demarque et al.\ (2004) and D'Antona \& Mazzitelli
 (1997): both predict ages slightly smaller than 1 Myr for HP Tau/G3,
 and somewhat larger than 5 Myr for HP Tau/G2. In principle,
 those differences could be real. In should be noticed, however, that 
 the vast majority of low-mass stars in Taurus (with spectral types M
 and late K) have ages smaller than 3 Myr (Brice\~no et al.\ 2002). 
 Moreover,  mass-dependent systematic effects in the age predictions 
 made by evolutionary tracks have been reported before. In
 particular, Hillenbrand et al.\ (2008) argued that existing
 models could significantly over-predict the age of  relatively 
 massive stars (M $\gtrsim$ 1.5 \Msun). HP Tau/G2 is precisely
 such a fairly massive star. So is HDE 283572, another young
 star in Taurus with a recently measured accurate distance (Torres
 et al.\ 2007).  The age estimate for that star based on the models 
 by Siess et al.\ (2000), Demarque et al.\ (2004) and D'Antona \& 
 Mazzitelli (1997) is 6--10 Myr (Fig.\ 7), somewhat larger than would 
 be expected for Taurus.
 The only of the four models considered here to predict similar ages
for the three members of the HP Tau group is that of Palla \& Stahler
(1999). Within the errors, all three stars fall on the 3 Myr isochrone. 
Note that this value is also consistent with the ages of lower mass
stars in Taurus (Brice\~no et al.\ 2002, see above).

\section{Conclusions and perspectives}

In this article, we have reported multi-epoch phase-referenced VLBA
observations of the weak-line T Tauri star HP Tau/G2 located near the
eastern edge of the Taurus star-forming complex. These observations
allowed us to measure the trigonometric parallax of the target with an
accuracy better than 1\%, and to refine the determination
of the intrinsic parameters of the source. Combined with previous
similar results on other young stars of Taurus, these data also
enabled us to probe directly for the first time the depth of this
important region of star-formation. We found that HP Tau/G2 is about
30 pc farther than two stars (Hubble 4 and HDE~283572) located close
to the dark cloud Lynds 1495, near the central portion of Taurus. This
implies that the Taurus complex is at least as deep as it is wide on the
plane of the sky.  The famous young star T Tauri, located to the south
of the complex happens to be at an intermediate distance.

Our observations also allow us to determine the full velocity vector
of our sources with excellent accuracy. Combining the results from the
four stars considered so far, we estimate the mean peculiar velocity
of Taurus to be about 10.6 km s$^{-1}$, oriented almost entirely along
the direction of the Galactic plane. The lack of a significant vertical component may appear 
somewhat surprising given the location of Taurus about 40 pc below
the Galactic mid-plane. This might suggest that Taurus has reached its 
farthest distance from the mid-plane and is about to fall back towards it.
Overall, the peculiar velocity of Taurus appears to be in reasonable
agreement with measurements of the velocity dispersion of giant
molecular clouds and young main sequence stars in the Solar
neighborhood. 

Using our improved distances, we have refined the determination 
of the location of the stars in the HP Tau group on an HR diagram,
and compared those positions with theoretical models available
in the literature. There is reasonable agreement (within 40\%) 
between the different models on the mass of the stars. It is noteworthy,
however, that this agreement becomes progressively poorer as one
considers less massive stars. Although the different members of
the HP Tau group might be expected to be coeval, three of the theoretical
models considered here predict significantly different ages for the
various members. Moreover, those models predict ages for the
most massive member of the group (HP Tau/G2) somewhat larger
than would expected for Taurus. The only model for which all three 
stars in the HP Tau group fall on a single isochrone (3 Myr) is
that of Palla \& Stahler (1999). Similar studies of multiple young stellar 
systems would clearly help test and improve pre main sequence
evolutionary models.

\acknowledgements
R.M.T., L.L. and L.F.R.\ acknowledge the financial support of DGAPA,
UNAM and CONACyT, M\'exico. We are indebted to Tom Dame for sending
us a digital version of the integrated CO(1-0) map of Taurus, to Andy Boden
for his help with the PMS models, and to Cesar Brice\~no for his detailed
comments of the errors affecting the determination of the luminosity and
effective temperature of the young stars in the HP Tau group. We are
also grateful to the anonymous referee, in particular for his/her comments 
which prompted us to expand significantly our discussion of the comparison 
with PMS models. The National Radio Astronomy Observatory is a facility of 
the National Science Foundation operated under cooperative agreement 
by Associated Universities, Inc.

\end{document}